\documentstyle[preprint,prd,aps]{revtex}

\tighten

\begin{document}

\preprint{TRI-PP-94-22 \ --- \
SFU HEP-114-94 \ --- \ hep-lat/9409010 \ --- \ April 1994}

\title{Exploring lattice quantum chromodynamics by cooling}

\author{Howard D. Trottier}
\address{Department of Physics, Simon Fraser University,
Burnaby, B.C. Canada V5A 1S6}

\author{R. M. Woloshyn}
\address{TRIUMF, Theory Group,
4004 Wesbrook Mall, Vancouver, B.C. Canada V6T 2A3}


\maketitle

\begin{abstract}
The effect of cooling on a number of observables is calculated
in SU(2) lattice gauge theory. The static quark-antiquark
potential and spin-dependent interactions
are studied, and the topological charge is monitored.
The chiral symmetry breaking order parameter
$\langle \overline{\chi}\chi \rangle$ and meson correlators
are calculated using staggered fermions.
Interactions on the distance scale of a few lattice
spacings are found to be essentially eliminated by cooling,
including the spin-dependent potentials.
$\langle \overline{\chi}\chi \rangle$
and meson correlators up to time separations of several
lattice spacings relax very quickly to their free-field values.
At larger times, there is evidence of a difference
between the pseudoscalar and vector channels.
A fit to the pseudoscalar correlation function yields
``mass'' values about $2/3$ (in lattice units)
of the uncooled masses.
These results raise the question of how to reconcile the
large-time behavior of the hadron correlators
with the fact that the spin-dependent potentials and
$\langle \overline{\chi}\chi \rangle$ essentially disappear
(in lattice units) after only a small amount of cooling.
\end{abstract}


\section{Introduction}

Cooling in lattice field theory is a technique for exposing the
topological features of field configurations \cite{berg,teper1}.
Recently, using this method, evidence has been
presented \cite{chu,chuetal} that instantons play a
dominant role in determining hadron properties in quantum
chromodynamics (QCD). This conclusion was based
on the calculation of long-distance properties of a
variety of hadronic correlation functions. It has been argued,
however, that the persistence of long-distance effects
(particularly, confinement \cite{campo}) is an inevitable
consequence of the local nature of the cooling procedure and is not
indicative of the underlying dynamics \cite{teper2}.

To gain some more insight into what cooling is doing
we have extended our work \cite{qed3} on three-dimensional QED to
four-dimensional QCD with SU(2) color.
We first examine the behavior of the static quark-antiquark potential
(calculated from Wilson loops) and of the spin-dependent interaction
(calculated from chromo-magnetic field correlations) under cooling.
As in QED$_3$, there is a good indication that the confining behavior
of the static potential persists even after a significant amount of
cooling although the overall scale of the potential is greatly reduced.
On the other hand, the spin-dependent potentials
(which are short-ranged) disappear rapidly upon cooling.
In the spin-spin interaction, a residual
effect, which one may associate with instantons, of only
a few percent of the orginal potential was observed.

Next chiral symmetry breaking was calculated using staggered fermions
in quenched approximation. The chiral symmetry order parameter
$\langle \overline{\chi} \chi \rangle$ was calculated at a few
nonzero quark mass values and compared to the result obtained for
free fermions. Our first observation is that
$\langle \overline{\chi} \chi \rangle$ relaxes to its free-field
value even after a small amount of cooling. This is an indication that
dynamical mass generation in the cooled configurations is
small compared to that
found in the uncooled vacuum. The extrapolation to
zero quark mass is problematic since, in principle, it requires
an extrapolation to infinite volume first. However, we can
conservatively estimate that in the region of intermediate cooling
(20 to 50 cooling steps) the chiral condensate has
been reduced by at least an order of magnitude.

Finally, meson correlators for the spin-0 and spin-1
channels were calculated.
The behavior of these time-correlation functions upon cooling
is in qualitative agreement with what would be
anticipated given the behavior
of the potentials and the chiral order parameter.
The meson correlators do not
relax uniformly. The relaxation rate
is most rapid at small times and it also decreases
as quark mass decreases.
At short time separations the
spin-0 and spin-1 meson correlators reflect very clearly the
effect of cooling on the spin dependent potential (degeneracy
of pseudoscalar and vector states) and on chiral symmetry breaking
(parity doubling in the spin-1 channel). At larger times at fit to the
pseudoscalar correlation functions yields ``mass'' values roughly 2/3
the size (in lattice units) of the uncooled masses.

Section II contains a description of the calculational methods
used in this paper. The results are presented in Sec. III.

\section{METHOD}

The usual plaquette action
\begin{eqnarray}
   S = \beta \sum_{x, \mu > \nu}
   \{ 1 - \frac{1}{2}\mbox{Tr} \, U_{\mu \nu}(x) \}
\end{eqnarray}
is used.
Periodic boundary conditions were
imposed on the gauge field links in all directions.
Field configurations were generated using a
heatbath Monte Carlo algorithm.

For cooling, links were updated ``vectorially'' following the same
checkerboard sequence as was used in the Monte Carlo algorithm.
Each link in turn was replaced by a link proportional to the
inverse of the sum of the ``staples'' of the plaquettes containing
the link being updated so as to minimize the local contribution
to the action.
Some test runs were done with adiabatic cooling \cite{campo2}.
After 10 to 20 cooling steps the results become
qualitatively the same as with full local minimization.

The physics of cooled configurations has been interpreted in terms
of instantons. We also monitor the topological charge  as a function
of cooling and, following Chu {\it et al.} \cite{chuetal}, the simple
transcription of $F_{\mu \nu}\tilde{F}_{\mu \nu}$ to the
lattice \cite{divec} is used for the topological charge density
\begin{equation}
   q_t(x) = {1 \over 32\pi^2} \sum_{\mu\nu\rho\sigma}
   \epsilon_{\mu\nu\rho\sigma} \mbox{Tr}
   \left\{ U_{\mu\nu}(x) U_{\rho\sigma}(x) \right\} ,
\end{equation}
where $U_{\mu \nu}$, $U_{\rho \sigma}$ are plaquettes of gauge
field links. The total topological charge is
\begin{equation}
   Q_t = \sum_x q_t(x) .
\end{equation}

As is well known this simple definition of topological charge
does not accurately represent the topological charge of the
uncooled lattice configurations \cite{lusch}. However, for
sufficiently smooth (cooled) configurations one can see
instantons with the expected continuum action.

The Creutz ratio
\begin{eqnarray}
   C(R,T) = -\ln \, \frac{W(R,T)\,W(R-1,T-1)}{W(R-1,T)\,W(R,T-1)} ,
\label{creutz}
\end{eqnarray}
where $W(R,T)$ denotes the $R$ by $T$ rectangular Wilson loop,
can be used to determine the string tension.
For large loops, which obey the area law, $C(R,R)$
gives the string tension. The potential $V(R)$ between
static quarks was also calculated directly by extrapolating
Wilson loops to large $T$
\begin{eqnarray}
V(R)= - {\lim_{T \to \infty}}
\,\, \frac{1}{T} \,\, \ln \,\, W(R,T).
\end{eqnarray}
Variance reduction methods \cite{VarRed} were used in
the computation of the Wilson loops in the uncooled theory.

In addition to the confining central potential,
spin-dependent interactions can also be calculated \cite{michael}.
These are related to chromo-magnetic field correlations
(see Ref. \cite{katz} for a simple derivation) which are computed by
making magnetic field insertions on $R \times T$ Wilson loops.
The spin-spin and tensor interactions are then given by
\begin{eqnarray}
   & & a^3 g^2 V_{SS}(R) = \nonumber \\
   & & {1 \over T-1}
   \sum_{t_1=T/2, T/2 \pm1} \,\, \sum_{t_2=1}^{T-1} \,\,
   { \langle B_L(0,t_1) B_L(R, t_2) \rangle
         + 2 \langle B_{\perp}(0,t_1) B_{\perp}(R, t_2) \rangle
   \over W(R,T) } ,
\end{eqnarray}
and
\begin{eqnarray}
   & & a^3 g^2 V_T(R) = \nonumber \\
   & & {1 \over T-1}
   \sum_{t_1=T/2, T/2 \pm1} \,\, \sum_{t_2=1}^{T-1} \,\,
   { \langle B_L(0,t_1) B_L(R, t_2) \rangle
         - \langle B_{\perp}(0,t_1) B_{\perp}(R, t_2) \rangle
   \over W(R,T) }
\end{eqnarray}
where $\langle B_L(0,t_1) B_L(R, t_2) \rangle$ and
$\langle B_{\perp}(0,t_1) B_{\perp}(R, t_2) \rangle$
are expectation values of Wilson loops with plaquette insertions
(at $t_1$ and $t_2$) corresponding to magnetic fields parallel and
transverse to the spatial direction of the loop.
In practise, the magnetic field insertion $B$ that was used was the
average over the eight spatial plaquettes whose corners
lie on the Wilson loop $W(R,T)$. This corresponds to
operator II of Ref. \cite{michael}.

In addition to observables constructed purely from gauge field
variables we are also interested in how quarks behave in
the cooled vacuum. In QCD a basic property of the
vacuum is chiral symmetry breaking
which can be studied most easily if staggered fermions are used.
The action for staggered fermions is
\begin{eqnarray}
   S_f & = & \frac{1}{2} \sum_{x,\mu} \eta_\mu(x)
   \left[ \overline{\chi}(x) \, U_\mu(x)
   \,\chi_{(x+\hat{\mu})} \,\, -\overline{\chi}_{(x+\hat{\mu})}
   U^{\dagger}_\mu(x) \, \chi(x) \right]
   + \sum_x \, m \overline{\chi}(x) \chi(x),
\nonumber \\
   & \equiv & \overline{\chi} \,\, {\cal M}(\{U\}) \, \chi  ,
\label{Sf}
\end{eqnarray}
where $\overline{\chi}, \chi$ are single-component fermion fields,
$\eta_\mu(x)$ is the staggered fermion phase \cite{kawo},
$m$ is the mass in lattice units and the $U's$ are gauge field links.
Antiperiodic boundary conditions were used for the fermion
fields in all directions.

The chiral symmetry order parameter is calculated from the
inverse of the fermion matrix ${\cal M}$ of Eq. (\ref{Sf})
\begin{eqnarray}
   \langle \overline{\chi} \chi \rangle
   = \frac{1}{V} \, \langle \mbox{Tr} \, {\cal M}^{-1}(\{U\}) \rangle ,
\end{eqnarray}
where V is the lattice volume and the angle brackets
denote the gauge field configuration average. A random source
method \cite{scalet,fiebig} was used to calculate
$\mbox{Tr} {\cal M}^{-1}(\{U\})$. Sixteen Gaussian random sources
were used for each gauge field configuration.

Meson correlation functions can be constructed from local bilinears
of the single-component $\chi$ fields. We consider two such correlators
which after integration over the fermion fields, can be expressed in
terms of the inverse of the fermion matrix as
\begin{equation}
   g_0(t) = \sum_{\vec x} \mbox{Tr} \left\{
   {\cal M}^{-1}(\vec x, t; 0)
   \left[ {\cal M}^{-1}(\vec x, t; 0) \right]^\dagger \right\},
\end{equation}
and
\begin{equation}
   g_1(t) = \sum_{\vec x}
   \left[ (-1)^{x_1} + (-1)^{x_2} + (-1)^{x_3} \right]
   \mbox{Tr} \left\{ {\cal M}^{-1}(\vec x, t; 0)
   \left[ {\cal M}^{-1}(\vec x, t; 0) \right]^\dagger  \right\}.
\end{equation}
These two functions describe the propagation of zero-momentum
meson states of spin 0 and 1 respectively. As is well known,
with local staggered fermion operators mixing between
states of different parity can in principle occur. In practise, the
spin-0 channel is essentially pure pseudoscalar and describes the
pseudo-goldstone boson. The spin-1 correlator is dominated by the
vector meson (at least in the uncooled theory) with some admixture
of axial-vector meson states.

\section{RESULTS}

Most of the calculations were done on a $12^4$
lattice at $\beta = 2.4$. This value was chosen as it
is well into the scaling region for SU(2) color. For comparison
some calculations were done at $\beta = 2.2$ on a $12^4$ lattice
and at $\beta = 2.4$ on a $16^4$ lattice. However,
not all results will be shown here since they are qualitatively
the same in all cases.

It is useful to consider first the topological properties of the gauge
field configurations. A sample of 300 configurations, separated by 100
heat-bath Monte-Carlo sweeeps after 4000 sweeps of thermalization, was
analyzed. In the uncooled configurations our value for the (lattice)
topological susceptibility
$\langle Q^2_t/L^4 \rangle$ of $(3.7 \pm 0.3)\times10^{-5}$
agrees well with the high statistics result of
$(3.5 \pm 0.1)\times10^{-5}$ obtained by
Campostrini {\it et al.} \cite{campo2}. The
simple transcription of the continuum topological charge operator
is not a true topological quantity on the lattice \cite{lusch}.
It need not take integer values as can be seen in Fig. 1
which shows a histogram of the number of configurations versus
$Q_t$. However in cooled configurations which are sufficiently
smooth, the operator $Q_t$ does cluster around integer values.
This is shown in Fig. 2 constructed from our sample of
300 configurations at 25, 50, 75 and 100 cooling steps.

It is also useful to examine the behavior of the action under cooling.
Histograms of the average plaquette are plotted in Fig. 3 for different
amounts of cooling. Fig. 4 shows a scatter plot of $Q_t$ versus average
plaquette. At 100 cooling steps there is a fairly obvious instanton
interpretation. The vertical dashed lines in Fig. 3d correspond to
values of the total action of $8 \pi^2n/g^2$ for n=1, 2, 3 and 4.
After about 100 cooling steps the configurations are dominated
by a single classical instanton. 75 cooling steps seems
to be in a transition region \cite{poli}. Single instanton peaks
are seen in the action but there are many configurations which have
a more complicated structure. The region of 25 to 50 cooling
has been interpreted as being dominated by
multi-instanton--anti-instanton fluctuations. However
one has to be aware that, as can be inferred from
Fig. 3 and 4, these configurations are not simply superpositions of
isolated (noninteracting) classical instantons.

The Creutz ratio and static potential can be calculated
from Wilson loops as in Eqs. (4) and (5).
A $12^4$ lattice is too small to see the true asymptotic
confining behavior but we can infer the general trend.
The Creutz ratio $C(R,R)$ calculated from our sample
of 300 uncooled configurations is shown in Fig. 5(a).
On the same graph we also show the Creutz ratio after 25 cooling
steps. Further cooling results in the Creutz ratios of Fig. 5(b).

The corresponding results for the potentials are shown in Fig. 6.
The potentials are steadily reduced by cooling but the
concave upward curvature is consistent with the idea
put forward by Teper \cite{teper2} that the string tension survives
cooling, albeit, at increasingly large distances.

Spin-dependent potentials are calculated from chromo-magnetic field
correlations. The spin-spin and tensor potentials are plotted in
Fig. 7 and 8 respectively. The uncooled potentials are qualitatively
consistent with the $\beta = 2.575$ operator II results of Michael
and Rakow \cite{michael}. The spin-dependent potentials are short
ranged and decrease rapidly upon cooling. In the region of 25 to 50
cooling steps the residual effect in these potentials is only
at the level of a few percent of the uncooled values.

Of course it has to be remembered that the comparison of cooled and
uncooled results is being done here in terms
of lattice (not ``physical'') units. It is clear that if one
attempted to keep the magnitude of the
potentials quantitatively similar in physical units while cooling
a very large change in the lattice spacing would be required.

Spontaneous chiral symmetry breaking is a basic property of QCD. A
signature for this phenomenon is the persistence of a nonzero value
of the ``quark condensate'' in the limit of zero quark mass. To study
chiral symmetry breaking we calculate the expectation value of the
local staggered fermion operator
$\langle \overline{\chi}\chi \rangle$. For our
$12^4$ calculation, 40 configurations (separated by 200 Monte Carlo
sweeps) at $\beta = 2.4$ and staggered fermion masses $ ma = 0.3,
0.2$ and $0.1$ were used. Fig. 9 shows
$\langle \overline{\chi}\chi \rangle$ versus
quark mass (in lattice units). The results without cooling are
consistent with those of Billoire {\it et al.} \cite{bill} at the same
value of $\beta$.
The quantity $\langle \overline{\chi}\chi \rangle$
was also calculated using 20 configurations on a $16^4$ lattice
and quark masses down to $ma=0.05$. The results are shown in Fig. 10.
At the masses that are common to both calculations, the
$12^4$ and $16^4$ lattice results are consistent.
The results for
$\langle \overline{\chi}\chi \rangle$ after 10 cooling steps
and for massive free staggered fremions are also plottted in
Figs. 9 and 10. After 10 cooling steps the values of
$\langle \overline{\chi}\chi \rangle$ are
already quite close to the free field results
at the same value of $ma$. Therefore, dynamical mass generation
seems to be quite small in the cooled vacuum.

In principle the extrapolation of $\langle \overline{\chi}\chi \rangle$
to zero quark mass to extract the genuine chiral condensate
requires that the infinite volume limit be taken first.
This would allow calculations at arbitrarily small quark masses.
In practise we can only do a limited number of calculations
so the procedure of Billoire {\it et al.} \cite{bill}
is adopted. Three values of
$\langle \overline{\chi}\chi \rangle$ at the lowest
nonzero mass are used to determine the coefficients of the expansion
\begin{equation}
   \langle \overline{\chi}\chi \rangle(m)
   = \langle \overline{\chi}\chi \rangle_0
   + \langle \overline{\chi}\chi \rangle_1 \, m
   + \langle \overline{\chi}\chi \rangle_2 \, m^2 .
\end{equation}

The extrapolated values of $\langle \overline{\chi}\chi \rangle_0$
from this procedure are plotted in Fig. 11 versus cooling
up to 50 cooling steps. Unfortunately the
values from the $12^4$ and $16^4$ lattice calculations do not
agree indicating that perhaps the correct mass
window is not being used for the extrapolation or
that the extrapolation function is not adequate.
Therefore only a qualitative conclusion is possible, namely, that
the chiral condensate (in lattice units) is greatly reduced in cooled
configurations. Of course, one could keep the condensate fixed in
physical units but this would require a decrease of the lattice
spacing by a factor of about 2.5 to 3.

Finally we examine the meson correlation functions.
First we consider the qualitative effects of cooling by comparing
the meson correlators in the vector and pseudoscalar channels. A
sample of results at the lightest quark mass values for both
$12^4$ and $16^4$ lattice are presented.

Fig. 12 shows the pseudoscalar and vector zero-momentum correlators
on the $12^4$ lattice for $ma = 0.1$.
After only a small amount of cooling the correlation functions
for small time separations display essentially the same
behavior as would be obtained for free fields.
The pronounced oscillation of the
vector correlator indicates near degeneracy of even and odd
parity spin-1 states (parity doubling). It is clear that,
after cooling, pseudoscalar and vector states are also nearly
degenerate. In other words, after 20 to 50 cooling steps, quark
propagation at $ma = 0.1$, at least in the limited time region
available on the $12^4$ lattice, shows very little evidence of
spin-dependent forces or chiral symmetry breaking.

On the $16^4$ lattice the correlations functions can be explored over
a larger time range and at smaller masses. The results for $ma = 0.1$
and $ma = 0.05$ are plotted in Figs. 13 and 14 respectively.
At large time separations the correlation functions differ from
those obtained for free fields. For example, comparing
Figs. 12 and 13 one sees at the largest time separation
some evidence of a difference between the pseudoscalar and
vector channels. Similarly, if the mass is decreased (compare
Figs. 13 and 14) the difference between pseudoscalar and vector
correlators after cooling is enhanced.

Although our lattices are too small to do very accurate mass
determinations we were able to get reasonable fits to the
pseudoscalar correlator on the  $16^4$ lattice in a restricted
time window with the function
\begin{equation}
   g_0(t) = C \left[ e^{-M_P t} + e^{-M_P (L-t)} \right],
\end{equation}
where $L$ is the size of the lattice.
The meson decay constant $f_P$ can be calculated \cite{Hamber}
from the coefficient $C$ according to
$f_P = 2 m \sqrt{C} / M_P^{3/2}$.
The results for the mass and decay constant are given in Table 1.
The pattern is consistent with the findings of
Chu {\it et al.} \cite{chuetal}. The masses (in lattice units)
are reduced to about 2/3 of the uncooled values and do
not change very much in the range of intermediate cooling
(20 to 50 cooling steps). The meson decay constant, which
reflects the short-distance behavior of the wave function,
is reduced to about 1/2 of the uncooled value.

How is one to interpret the behavior of the meson
correlators under cooling? One way is that the correlators
reflect the true dynamics of the cooled vacuum. At sufficiently
large times and sufficiently small masses the
correlation functions are not affected by
cooling very much so hadron properties (and the uncooled
lattice scale) survive essentially intact.
This is the approach of Chu {\it et al.} \cite{chuetal}.
An obvious question with this interpretation
is how to reconcile it with the finding of direct calculation that
the spin-dependent potentials and chiral symmetry breaking essentially
disappear (in lattice units) after only a small amount of cooling.

Another interpretation is that the effect identified
by Teper \cite{teper2} in the string tension is also present in
the meson correlators. Namely, correlations functions do not
relax uniformly under cooling. At sufficiently large time
separations it is inevitable that the behavior of the original
uncooled configurations persists and this does not reflect
the dynamics of the cooled vacuum.

\section{SUMMARY}

In this work the effect of cooling on a
number of observables was calculated in SU(2) lattice gauge
theory. These include the central and spin-dependent potentials, the
chiral symmetry breaking order parameter and meson correlators.

Even after 100 cooling steps a remnant of the confining static
potential is seen at large distance. However interactions on
the distance scale of a few lattice spacings are essentially
eliminated. This includes the spin-dependent interactions induced
by chromo-magnetic field correlations.

The quantity $\langle \overline{\chi}\chi \rangle$ was found to
approach its free-field value very quickly with cooling. This
may indicate that on the lattice, with staggered fermions, chiral
symmetry breaking is driven more by local fluctuations in the
topological charge density rather than by global topological
properties.  As cooling smooths out local fluctuations,
chiral symmetry breaking decreases rapidly, even though
instantons remain.  This is in line with the study of
Hands and Teper \cite{hands} who suggest that the instanton-induced
zero modes for staggered fermions do not become delocalized
as would be required for chiral symmetry breaking.

Meson correlators up to time separations of several
lattice spacings relax quickly to free-field values reflecting
the behavior of the potentials. At larger times, differences
from free-field behavior persist.
A fit to the pseudoscalar
correlation function yields ``mass'' values about $2/3$ of
the uncooled masses. To this extent our results confirm
the calculation of Chu {\it et al.} \cite{chuetal}.

On the other hand, we cannot conclude that our results
provide evidence for the dominant role of instantons.
There are properties of QCD that are changed by cooling.
The lack of a direct signature for spin-dependent forces and
chiral symmetry breaking has to be reconciled with the large-time
behavior of the hadron correlators. Until this can be done the
interpretation of the hadron properties after cooling
remains imprecise.

\acknowledgments
It is a pleasure to thank M.-C. Chu, G. Poulis and K. Yee
for helpful discussions.
This work was supported in part by the Natural Sciences and
Engineering Research Council of Canada.

\newpage


\begin{figure}
\caption{Histogram of the number of configurations versus the absolute
value of the topological charge in configurations with no cooling.}
\end{figure}

\begin{figure}
\caption{Histogram of the number of configurations versus the absolute
value of the topological charge after (a) 25 cooling steps,
(b) 50 cooling steps, (c) 75 cooling steps, and (d) 100 cooling steps.}
\end{figure}

\begin{figure}
\caption{Histogram of the number of configurations versus the average
plaquette after (a) 25 cooling steps,
(b) 50 cooling steps, (c) 75 cooling steps, and (d) 100 cooling steps.
The vertical dashed lines in (d) indicate the value of the average
plaquette corresponding to an instanton action with
winding number 1,2,3 and 4.}
\end{figure}

\begin{figure}
\caption{Scatterplot of the topological charge versus the average
plaquette after (a) 25 cooling steps, (b) 50 cooling steps,
(c) 75 cooling steps, and (d) 100 cooling steps.}
\end{figure}

\begin{figure}
\caption{The Creutz $C(R,R)$ as a function of loop size $R$ for
(a) no cooling  ($\triangle$) and 25 cooling steps ($\bullet$),
(b) 25 cooling steps ($\bullet$), 50 cooling steps
(\protect{\rule{2mm}{2mm}})
and 100 cooling steps (filled triangles).}
\end{figure}

\begin{figure}
\caption{The static potential $V(R)$ versus $R$ for
(a) no cooling ($\triangle$) and 25 cooling steps ($\bullet$),
(b) 25 cooling steps ($\bullet$), 50 cooling steps
(\protect{\rule{2mm}{2mm}})
and 100 cooling steps (filled triangles).}
\end{figure}

\begin{figure}
\caption{The spin-spin potential $V_{SS}(R)$ versus $R$ for
(a) no cooling  ($\triangle$) and 25 cooling steps ($\bullet$),
(b) 25 cooling steps ($\bullet$), 50 cooling steps
(\protect{\rule{2mm}{2mm}})
and 100 cooling steps (filled triangles).}
\end{figure}

\begin{figure}
\caption{The tensor potential $V_T(R)$ versus $R$ for
(a) no cooling  ($\triangle$) and 25 cooling steps ($\bullet$),
(b) 25 cooling steps ($\bullet$), 50 cooling steps
(\protect{\rule{2mm}{2mm}})
and 100 cooling steps (filled triangles).}
\end{figure}

\begin{figure}
\caption{The chiral order parameter $\langle \overline{\chi}\chi \rangle$
versus fermion mass $ma$ calculated on a $12^4$ lattice
for no cooling ($\triangle$) and 10 cooling steps ($\Box$).
The lattice free-field values are also shown ($\circ$).}
\end{figure}

\begin{figure}
\caption{The chiral order parameter $\langle \overline{\chi}\chi \rangle$
versus fermion mass $ma$ calculated on a $16^4$ lattice
for no cooling ($\triangle$) and 10 cooling steps ($\Box$).
The lattice free-field values are also shown ($\circ$).}
\end{figure}

\begin{figure}
\caption{The chiral order parameter $\langle \overline{\chi}\chi \rangle$
extrapolated to zero fermion mass as a function of cooling step
on the $12^4$ lattice ($\triangle$) and the
$16^4$ lattice ($\circ$).}
\end{figure}

\begin{figure}
\caption{Comparison of pseudoscalar ($\triangle$)
and vector ($\circ$) meson
correlators calculated on a $12^4$ lattice at $ma = 0.1$ for
(a) no cooling, (b) 10 cooling steps,
(c) 20 cooling steps, and (d) 50 cooling steps.}
\end{figure}

\begin{figure}
\caption{Comparison of pseudoscalar ($\triangle$)
and vector ($\circ$) meson correlators
calculated on a $16^4$ lattice at $ma = 0.1$ for
(a) no cooling, (b) 10 cooling steps,
(c) 20 cooling steps, and (d) 50 cooling steps.}
\end{figure}

\begin{figure}
\caption{Comparison of pseudoscalar ($\triangle$)
and vector ($\circ$) meson correlators
calculated on a $16^4$ lattice at $ma = 0.05$ for
(a) no cooling, (b) 10 cooling steps,
(c) 20 cooling steps, and (d) 50 cooling steps.}
\end{figure}

\begin{table}
\caption{Results of fits to pseudoscalar meson mass and decay
constant. A rough estimate of the systematic error due to
the fitting procedure is about 5\% in the values for
$M_P$, and about 10\% in the values for $f_P$.}

\begin{tabular}%
{c@{\hspace{20pt}}c@{\hspace{20pt}}c@{\hspace{20pt}}c@{\hspace{50pt}}%
 c@{\hspace{20pt}}c@{\hspace{20pt}}c@{\hspace{20pt}}c}

\\[-6pt]

  $ma$ & cooling & $M_P a$ & $f_P a$
& $ma$ & cooling & $M_P a$ & $f_P a$ \\
       &  step   &           &
&      &  step   &           &           \\ \hline\hline

\\[-6pt]

  0.05  &  0  &  0.61  &  0.13

& 0.10  &  0  &  0.83  &  0.17 \\

        & 10  &  0.45  &  0.077

&       & 10  &  0.56  &  0.089 \\

        & 20  &  0.38  &  0.075

&       & 20  &  0.58  &  0.085 \\

        & 50  &  0.34  &  0.070

&       & 50  &  0.51  &  0.075 \\[6 pt]

  0.20  &  0  &  1.11  &  0.22

& 0.30  &  0  &  1.33  &  0.26 \\

        & 10  &  0.74  &  0.11

&       & 10  &  0.98  &  0.13 \\

        & 20  &  0.74  &  0.10

&       & 20  &  0.94  &  0.12 \\

        & 50  &  0.74  &  0.10

&       & 50  &  0.92  &  0.11 \\ \hline

\end{tabular}
\label{CoolMpi}
\end{table}

\end{document}